\documentclass[12pt,reprint,tightenlines,groupedaddress,superscriptaddress,floatfix,aps]{revtex4}
\usepackage{graphicx}
\usepackage{rotating}
\usepackage{array}
\usepackage{multirow}
\usepackage{epstopdf}
\usepackage{amsmath}
\usepackage{amsfonts}
\usepackage{setspace}
\usepackage{braket}
\usepackage{color}

\usepackage{nameref,zref-xr} 
\zxrsetup{toltxlabel} 

\usepackage{hyperref}

\DeclareMathOperator{\Tr}{Tr}

\begin{document}
\title{Iterative subspace algorithms for finite-temperature solution of Dyson equation}
\author{Pavel Pokhilko}
\affiliation{Department  of  Chemistry,  University  of  Michigan,  Ann  Arbor,  Michigan  48109,  USA}
\author{Chia-Nan Yeh}
\affiliation{Department of Physics, University of Michigan, Ann Arbor, Michigan 48109, USA }
\author{Dominika Zgid}
\affiliation{Department  of  Chemistry,  University  of  Michigan,  Ann  Arbor,  Michigan  48109,  USA}
\affiliation{Department of Physics, University of Michigan, Ann Arbor, Michigan 48109, USA }

\renewcommand{\baselinestretch}{1.0}
\begin{abstract}
One-particle Green's functions obtained from the self-consistent solution of the Dyson equation 
can be employed in evaluation of spectroscopic and thermodynamic properties for both molecules and solids. 
However, typical acceleration techniques used in the traditional quantum chemistry self-consistent algorithms cannot be easily deployed for the Green's function methods, 
because of non-convex grand potential functional and non-idempotent density matrix. 
Moreover, the inclusion of correlation effects in the form of the self-energy matrix 
and changing chemical potential or fluctuations in the number of particles can make the optimization problem more difficult. 
In this paper, we study acceleration techniques to target the self-consistent solution of the Dyson equation directly.  
We use the direct inversion in the iterative subspace (DIIS), 
the least-squared commutator in the iterative subspace (LCIIS), 
and the Krylov space accelerated inexact Newton method (KAIN). 
We observe that the definition of the residual has a significant impact on the convergence of the iterative procedure.
Based on the Dyson equation, we generalize the concept of the commutator residual used in DIIS (CDIIS) and LCIIS, 
and compare it with the difference residual used in DIIS and KAIN. 
The commutator residuals outperform the difference residuals for all considered molecular and solid systems within both
GW and GF2.  
The generalized CDIIS and LCIIS methods successfully converged restricted GF2 calculations for a number of strongly correlated systems, which could not be converged before.  
We also provide practical recommendations to guide convergence in such pathological cases. 
\end{abstract}
\maketitle

\section{Introduction}
Self-consistent approximations to the Dyson equation are important for description of 
molecular and solid-state phenomena, 
such as photoionization\cite{Almbladh:photoemission:1985,Hedin:photoemission:1985,Fujikawa:photoelectron:chapter:2015}, thermodynamic properties\cite{Stefanucci:vanLeeuwen:book:2013,Fetter:Walecka:2012}, magnetic\cite{Pokhilko:local_correlators:2021} and non-equilibrium phenomena\cite{Stefanucci:vanLeeuwen:book:2013}. 
A full self-consistency guarantees several important theoretical features of the approximations, 
such as thermodynamic consistency, gauge invariance, and current continuity\cite{Baym61,Baym62}.  These theoretical features ensure that in practical calculations, the values of thermodynamic observable evaluated either using thermodynamic integration or employing the grand canonical partition function remain mutually consistent.

To access thermodynamic properties, we use the Matsubara imaginary time contour 
and define the imaginary-time one-particle Green's function $G$\cite{Mahan00,Negele:Orland:book:2018} as
\begin{gather}
G_{pq} (\tau) = -\frac{1}{Z} \Tr \left[e^{-(\beta-\tau)(\hat{H}-\mu \hat{N})} p e^{-\tau(\hat{H}-\mu \hat{N})} q^\dagger  \right], \\
Z = \Tr \left[ e^{-\beta(\hat{H}-\mu \hat{N})} \right],
\end{gather}
where $Z$ is the grand canonical partition function, 
$\beta$ is the inverse temperature, 
$\mu$ is the chemical potential, 
$\hat{N}$ is the particle-number operator, 
and $\tau$ is the imaginary time (usually defined as a series of grid points). 
In the molecular case, 
the indices $p,q$ enumerate spin-orbitals. 
In the solid case, 
the multiindices $p,q$ contain both enumeration of spin-orbitals and momentum $k$ 
(because of momentum conservation, only one index $k$ needs to be stored for the Green's function and self-energy).
In the imaginary frequency domain, the Dyson equation is expressed as
\begin{gather}
G^{-1}(i\omega_n) = G^{-1}_0(i\omega_n) - \Sigma[G](i\omega_n), \protect\label{eq:Dyson}
\end{gather}
where $\omega_n$ is the fermionic (odd) Matsubara imaginary frequency
\begin{gather}
\omega_n = \frac{\pi (2n+1)}{\beta},
\end{gather}
that is discrete (however, for the sake of brevity in the remaining text we list it as $\omega$ leaving the subscript $n$ denoting the grid points out).
$G_0$ is the zeroth order Green's function and $\Sigma[G](i\omega_n)$ is the self-energy functional depending on the full one-electron Green's function. 
The self-energy decomposes into the static and dynamical parts
\begin{gather}
\Sigma[G](i\omega) = F[G] + \Sigma^{dyn}[G](i\omega),
\end{gather}
where the static part is the Fock matrix built from the correlated one-electron density 
$\gamma=-G(\tau=\beta^-)$ in spin-orbitals. 
The dynamical part of the self-energy encodes the many-body correlational effects that are not described at the one-body level of theory.
While Eq.~\ref{eq:Dyson} is exact, in practice diagrammatic approximations to the self-energy are used. 
The specific approximations, such as GF2\cite{Snijders:GF2:1990,Dahlen05,Phillips14,Rusakov16,Welden16} 
and GW\cite{Hedin65,G0W0_Pickett84,G0W0_Hybertsen86,GW_Aryasetiawan98,Stan06,Koval14,scGW_Andrey09,GW100,Holm98,QPGW_Schilfgaarde}, define the explicit functional form of the self-energy. 

Due to the explicit dependence of the self-energy on the Green's function, the Dyson equation is solved iteratively. 
Unfortunately, in practice such an iterative procedure is often difficult to converge. 
The lowest-order approximation to the self-energy is static. It is called the Hartree--Fock (HF) approximation, 
which is a basic building block of many quantum chemical methods. 
In the zero-temperature formulation, 
the simplest Roothaan's direct steps\cite{Roothaan:HF:1951} for HF take the density matrix $\gamma$ from the current iteration, 
construct the new Fock matrix $F[\gamma]$, 
diagonalize it, and
assign its lowest eigenvectors as the occupied orbitals based on the Aufbau principle.
Subsequently, using these eigenvectors one generates a new density matrix. 
Such direct steps often converge slowly or even diverge, especially when a poor initial guess is used.  
To stabilize the convergence, a constant mixing with the previous iteration (so called ``damping'') is introduced 
for the Fock or density matrix \cite{Szabo_ostlund,Wang:damping:2001}
\begin{gather}
F_i^{mix} = \alpha F_i + (1-\alpha) F_{i-1}.
\end{gather}
This technique is commonly applied in the Green's function methods, where the full self-energy containing both static and dynamic parts instead of the Fock matrix is used for mixing.
However, in most cases, both in traditional HF and Green's function calculations, damping still leads to a relatively slow convergence of the iterative procedure.

A variety of techniques for acceleration and stabilization of the zero-temperature HF convergence has been developed, 
such as direct inversion in the iterative subspace\cite{Pulay:DIIS:1980,Pulay:82:DIISSCF} (DIIS), 
relaxed constrained algorithm\cite{Cances:RCA:2000} (RCA), 
energy-DIIS\cite{Kudin:EDIIS:2002}, 
ADIIS\cite{Yang:ADIIS:2010}, 
least-squares commutator in the iterative subspace\cite{LCIIS:2016} (LCIIS),
second-order methods\cite{Levy:SOSCF:1980,Chaban:SOSCF:1997,Neese:SOSCF:2000,Head-Gordon:GDM:2002}, 
linear-expansion shooting techniques\cite{Chen:LIST:2011,Wang:LISTb:2011} (LIST), 
maximum overlap methods\cite{Gill:MOM:08} (MOM). 
Since the initial guess has a big impact on the behavior of iterations, 
a number of advanced guess generators has been introduced\cite{Almof:SCF:1982,Lenthe:SAD:2006,Lehtola:SAP:2019}.
Calculations of periodic problems provide additional convergence challenges related to the singular behavior of Coulomb interaction, 
which is addressed by re-weighting contributions 
for each k-point and preconditioners\cite{Kerker:precond:1981,Kresse:DFT:algorithms:1996,Maschio:solid:DIIS:2018}.

Orbital optimization can be also done with the presence of electronic correlation. 
In wave-function methods, 
such as multi-configurational self-consistent field method\cite{OlsenText}, 
orbital-optimized coupled-cluster methods\cite{Handy:89:BD,ooccd,vooccd}, 
orbital-optimized M{\"o}ller--Plesset method\cite{MHG:O2:07,Neese:OO-SCS-MP2:2009,Bozkaya:OO-CCD:OO-MP2:2011}, the orbital optimization is very challenging and
convergence difficulties often limit applicability of these methods, 
which resulted in a development of advanced algorithms targeting this problem\cite{Yeager:MC-HF:1979,Roos:NR:CASSCF:1981,Werner:MCSCF:alg:I:2019,Werner:MCSCF:alg:II:2020}. 

\begin{figure}
  \includegraphics[width=7cm]{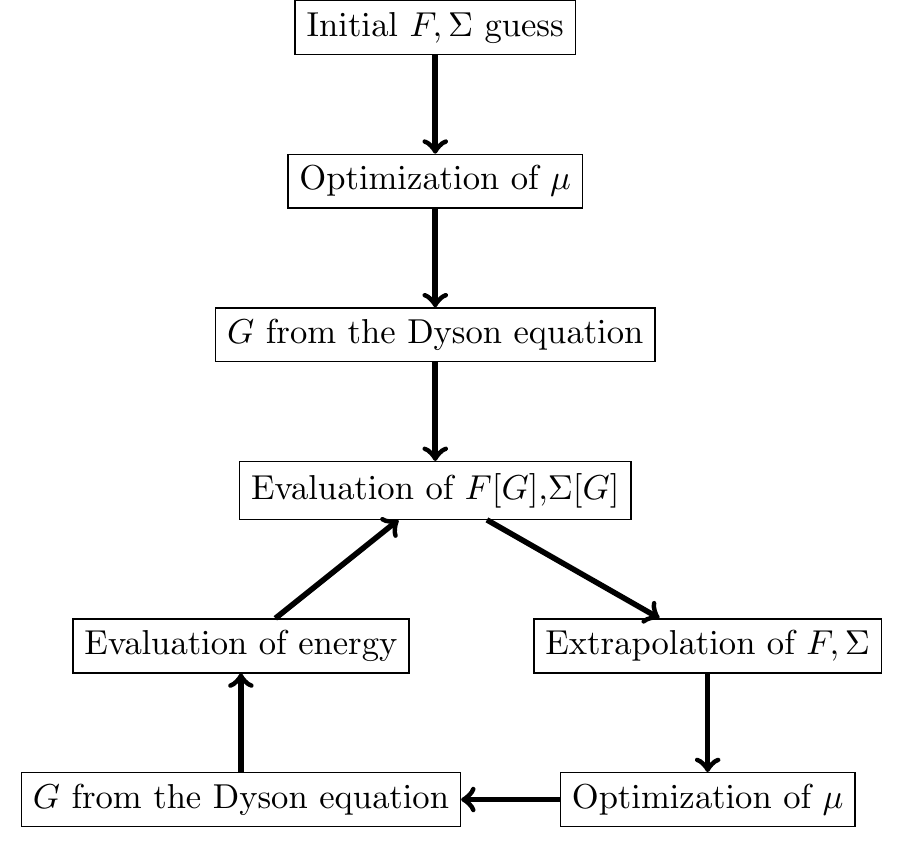}
\centering
\caption{A self-consistent scheme used in this paper. 
         The optimization of the chemical potential is needed when the Green's function is 
         required to reproduce a number of electrons, specified by user. 
         The extrapolation step depends on the algorithm used and is the object of study of this paper. 
         The initial approximation to the Fock matrix comes from 
         the converged zero-temperature calculation unless written explicitly in the text. 
         The self-consistency cycle is terminated, when the changes in energy, 
         chemical potential, and density matrices are smaller than the user-specified thresholds.
         \protect\label{fig:algs}}
\end{figure}
Green's function methods also encounter many of these challenges. 
Moreover, due to the nature of the Green's functions, 
many algorithms, used in the wave function domain, are not easily and directly applicable. 
For example, in the full self-consistent Green's function evaluation, 
the Fock matrix $F[\gamma]$ 
that depends on the full one-particle density matrix is updated 
together with the self-energy $\Sigma[G](i\omega)$. 
One of the consequences of the full self-consistency is the gauge invariance, 
which can be seen as orbital optimization in the presence of correlation.  
However, since the Green's function formulation does not rely on the explicit description of the orbital rotations, 
one cannot take advantage of orbital rotations in the exponential form with the second-order methods. 
There are also other differences. 
For instance, the non-convexity of the grand potential with respect to the Green's function\cite{Potthoff:density_matrix:2013,Kotliar:effective_action:2001,Tremblay:convexity:VCA:2008} 
violates initial assumptions of the RCA, EDIIS, and ADIIS algorithms, 
making their Green's function generalizations not mathematically rigorous.  
Moreover, at a finite temperature, the distinction between occupied and virtual orbitals is blurred, and 
the MOM method is not applicable in such cases. 
Additionally, since in the grand canonical ensemble, the calculations can be conducted for a fixed chemical potential $\mu$, the number of particles between iterations can fluctuate thus exacerbating the convergence problems.

Since many of the methods used for large realistic systems (such as G0W0, GW0, and quasiparticle GW)~\cite{Rinke_GW_review} either did not require self-consistency or relied only on a partial self-consistency, 
the construction of  efficient algorithms for the Green's function convergence received little attention previously. 
While for some model systems Green's function methods 
were converged fully self-consistently,  
these optimizations usually were performed in a small orbital space and in a narrow energy window, 
which either hid convergence problems in less trivial situations or 
were cheap enough that the acceleration was not necessary. 
The goal of this paper is to study applicability of the subspace extrapolation techniques to the 
solution of the finite-temperature Dyson equation. 
We formulate and compare DIIS (with commutator and difference residuals), LCIIS, 
and KAIN algorithms for Green's functions and apply them to both molecules and solids. 
A general self-consistency scheme used in this paper is shown in  Fig.~\ref{fig:algs}.

\section{DIIS}
One of the most successful techniques is DIIS extrapolation\cite{Pulay:DIIS:1980,Pulay:82:DIISSCF}. 
The formulation of DIIS is general, which enabled its use in 
a variety of contexts\cite{Pulay:DIIS:1980,Pulay:82:DIISSCF,Pulay:GDIIS:1984,Scuseria:86:DIIS,Karadakov:DIIS:SCF-MI:2001,Daul:DerDIIS:CPSCF:2003,Nanda:2PA:14,Nanda:EOMCCSD_StaticPol:2016,Nanda:RIXS:19,Kowalski:DIIS:CCGF:2018,Visscher:qpGW:scaling:2021}.
DIIS is a subspace method that uses memory of the previous iterations to provide a guess for the new iteration in the following way. 
Here, the vector quantities are labeled as $\mathbf{v}_i$, where the subscript enumerates iterations. 
The vectors can be represented as 
\begin{gather}
\mathbf{v}_i = \mathbf{v}_* + \mathbf{e}_i,
\end{gather}
where the final solution has a star subscript and the error vector is $\mathbf{e}_i$. 
The goal is to have the norm of the error vector decreased with iterations. 
DIIS does that via linear extrapolation 
\begin{gather}
\mathbf{v}_{\text{extr}} = \sum_i c_i \mathbf{v}_i = \mathbf{v}_* \sum_i c_i + \sum_i c_i \mathbf{e}_i, \\
\mathbf{e}_{\text{extr}} = \sum_i c_i \mathbf{e}_i,
\end{gather}
where $c_i$ are the extrapolation coefficients. 
In order to converge to the final solution, it is natural to use the constraint 
\begin{gather}
\sum_i c_i = 1. \protect\label{eq:DIIS_constr}
\end{gather}
DIIS seeks to minimize the Euclidean norm of the extrapolated error 
$||\mathbf{e}_{\text{extr}}||$ to determine the extrapolation coefficients.
The corresponding Lagrangian for minimization of the squared norm is 
\begin{gather}
L^\text{DIIS}(c_i, \lambda) = \frac{1}{2}\sum_{ij} c_i B_{ij} c_j - \lambda (1-\sum_i c_i), \\
B_{ij} = \braket{\mathbf{e}_i, \mathbf{e}_j}. 
\end{gather}
Assuming that the extrapolation coefficients are real, the differentiation gives 
a necessary condition for minimum
\begin{gather}
\begin{pmatrix}
\mathrm{Re}(B_{11}) & \cdots & \mathrm{Re}(B_{1n}) & 1 \\
\cdots & \cdots & \cdots \\
\mathrm{Re}(B_{1n}) & \cdots & \mathrm{Re}(B_{nn}) & 1 \\
      1    &  \cdots & 1 & 0
\end{pmatrix}
\begin{pmatrix}
c_1 \\
\cdots \\
c_n \\
\lambda
\end{pmatrix} = 
\begin{pmatrix}
0 \\
\cdots \\
0 \\
1
\end{pmatrix}. 
\end{gather}
However, the numerical solution of the equation system above could be plagued by a bad condition number. 
As the residuals become smaller, their overlaps become smaller, leading to the appearance of small numbers in the $B$ matrix, which worsens the condition number. 
Many practical implementations of DIIS apply tricks to mitigate this problem. 
As a simple preconditioning one can partition the equation system 
\begin{gather}
B \vec{c} - \lambda \vec{1} = 0.
\end{gather}
This equation system can be solved with $\lambda = 1$, which gives the correct set of $c_i$ up to a constant multiplier. 
Then the coefficients are scaled to produce the DIIS constraint, which gives the solution of the original minimization condition.

Of course, in practice the final solution $\mathbf{v}_*$ is not known. 
The exact errors $\mathbf{e}_i$ are replaced with the error estimates, 
which may depend on the nature of the $\mathbf{v}_i$ vectors. 
A common and universal strategy is to take the difference between the subsequent iterations, 
$\mathbf{e}_i = \mathbf{v}_i - \mathbf{v}_{i-1}$, as an error vector. 
This strategy has been successfully applied to coupled-cluster equations\cite{Scuseria:86:DIIS} and 
response coupled-cluster equations\cite{Nanda:2PA:14,Nanda:EOMCCSD_StaticPol:2016,Nanda:RIXS:19,Kowalski:DIIS:CCGF:2018}.  
The DIIS has been recently applied to quasiparticle self-consistent GW\cite{Visscher:qpGW:scaling:2021} (qsGW, not to be confused with the fully self-consistent GW used in this paper), 
where self-energy was extrapolated and the difference between the density matrices was used as residuals. 
Another variant common in SCF calculations (CDIIS\cite{Pulay:82:DIISSCF}) 
uses Fock matrices as $\mathbf{v}_i$ vectors and 
a commutator residual as error vectors
\begin{gather}
\mathbf{e}_i = \left[ F_i, P_i \right],
\end{gather}
where the $F_i$ are the Fock matrices and $P_i$ are the one-electron densities. 
Here, we write the commutator assuming that either the orbital set is orthonormal or that the overlaps have already been absorbed in the density matrix. 

For Green's function calculations, we use the total self-energy $\Sigma_i$ as the $\mathbf{v}_i$ vectors. 
\begin{gather}
\mathbf{v}_i = \begin{pmatrix}
 F_i \\
\Sigma_i^{dyn}(\tau)
\end{pmatrix}.
\end{gather}

We use both the difference residuals between the subsequent iterations and the generalized commutator residual, which follows from multiplication of both sides of the Dyson equation (Eq.~\ref{eq:Dyson}) by $G(i\omega)$ from the left and right and taking a difference at the same iteration $j$:
\begin{gather}
C_{jj}(i\omega) = \left[ G_j(i\omega), G_0^{-1}(i\omega) - \Sigma_j(i\omega) \right]. 
\end{gather}
We then evaluate the commutator-error overlaps and norms in the time domain integrating over time. 

If the Green's function is expected to reproduce a user-specified number of electrons, 
the update of the chemical potential changes the $G_0$. 
If the chemical potential changes, one may choose to re-evaluate the commutators or skip updating them.

We will comment now on the computational cost necessary to perform the DIIS procedure. 
For periodic systems, irrespective of the method used for evaluating the self-energy, the computational scaling of commutator evaluation is $O(n_\omega n_k n^3)$, 
where $n_\omega$ is the number of the frequency points, 
$n_k$ is the number of $k$-points, 
$n$ is the number of orbitals in the unit cell. 
This scaling is up to a prefactor the same as the evaluation of the Green's function from the Dyson equation. 
The cost of the Fourier transform of the commutator on the non-uniform grid is $O(n_\omega^2 n_k n^2)$.  
The overlap between the residuals is taken as a matrix trace, which is done as $O(n_\omega n_k n^2)$.  
Since for solids the Green's function and self-energy may take a sizeable amount of memory, 
in our implementation, we store the subspaces of residuals and self-energies 
on disk and read them as they need to be accessed. 
Note that this additional computational cost necessary to perform DIIS is modest in comparison to the cost of evaluation of self-energy in both GW and GF2 procedures 
($O(n_\omega n_{aux}^2n^2+ n_\omega n_{aux}n^3)$ for molecular GW and $O(n_\omega n_{aux}^2n^3)$ for molecular GF2, 
where $n_{aux}$ is the number of auxilliary basis functions in resolution-of-identity approximation). 

\section{LCIIS}
In the context of HF and DFT methods, Li and Yaron introduced LCIIS\cite{LCIIS:2016}, which finds a set of coefficients extrapolating both the Fock matrix and the density 
\begin{gather}
F_{\text{extr}} = \sum_i c_i F_i, \\
P_{\text{extr}} = \sum_i c_i P_i.
\end{gather}
Indeed, since the Fock matrix depends on the density linearly, one can expect that the extrapolation coefficients for both of them are the same\cite{note:DFT_lin}. 
Once the coefficients are found and the extrapolation is performed for the Fock matrix, the new density $P_{n+1}$ is found from the $F_{\text{extr}}$. 

The LCIIS extrapolation coefficients are found from the minimization condition of the objective function---the squared norm of the commutator
\begin{gather}
f(c) = ||[P_{\text{extr}}, F_{\text{extr}}]||^2
= \Tr [P_{\text{extr}}, F_{\text{extr}}]^\dagger [P_{\text{extr}}, F_{\text{extr}}].
\end{gather}
To perform the calculation of the objective function, a 4-index tensor is introduced 
\begin{gather}
T_{ijkl} = \Tr [P_i, F_j]^\dagger [P_k, F_l], \\
f(c) = \sum_{ijkl} c_i c_j c_k c_l T_{ijkl}. \protect\label{eq:lciis_obj_f}
\end{gather}
The objective function and its derivatives are easily written as polynomials depending on coefficients and $T_{ijkl}$\cite{LCIIS:2016}.
The objective function is minimized with the DIIS constraint on the coefficients (Eq.~\ref{eq:DIIS_constr}).
Using Lagrange multipliers, the minimum could be found by applying the Newton method:
\begin{gather}
L^\text{LCIIS}(c_i, \lambda) = f(c) - \lambda (1-\sum_i c_i), \\
\begin{pmatrix}
\mathrm{Re}(H_{11}) & \cdots & \mathrm{Re}(H_{1n}) & 1 \\
\cdots & \cdots & \cdots \\
\mathrm{Re}(H_{1n}) & \cdots & \mathrm{Re}(H_{nn}) & 1 \\
      1    &  \cdots & 1 & 0
\end{pmatrix}
\begin{pmatrix}
\Delta c_1 \\
\cdots \\
\Delta c_n \\
\lambda
\end{pmatrix}
= -
\begin{pmatrix}
g_1 \\
\cdots \\
g_n \\
0
\end{pmatrix}, \\
\vec{c}_\text{new} = \vec{c} + \alpha \Delta \vec{c}, 
\end{gather}
where $H$ is the Hessian of the objective function $f(c)$, $g$ is the gradient of the objective function, 
and $\Delta c$ is the Newton step direction for the coefficients. 
The same partitioning of the system as for DIIS could be used to improve the condition number. 
In our implementation, we use a modified backtracking line search for the step size $\alpha$, 
where instead of a gradient of a function, only its tangential component to the constraint manifold is taken. 
This ensures that the Newton step does not violate the DIIS constraint (Eq. ~\ref{eq:DIIS_constr}).
The minimization of the objective function continues until convergence is achieved. 

However, it has been found that LCIIS also works for DFT\cite{LCIIS:2016}, where the exchange-correlation potential does not depends linearly on the density. A generalization of the LCIIS algorithm to the Green's functions is straightforward. 
Introducing extrapolations for the Green's function and the self-energy as well as the commutator
\begin{gather}
\Sigma_{\text{extr}} = \sum_i c_i \Sigma_i, \\
G_{\text{extr}} = \sum_i c_i G_i, \\
C_{ij}(i\omega) = \left[ G_i(i\omega), G_0^{-1}(i\omega) - \Sigma_j(i\omega) \right],
\end{gather}
the LCIIS algorithm can be applied to the solution of the Dyson equation. 
The cost of the commutator evaluation is the same as in the CDIIS generalization. 
The only substantial cost difference is due to the number of necessary commutators that
grows as $N^4$ with respect to the subspace size. 
However, in all practical cases studied here, this overhead is very small. 

One can suggest a modification to LCIIS procedure (called here modLCIIS), where 
the extrapolation coefficients for the Green's function and for the self-energy are different:
\begin{gather}
\Sigma_{\text{extr}} = \sum_i c_i^\Sigma \Sigma_i, \\
G_{\text{extr}} = \sum_i c_i^G G_i. 
\end{gather}
The minimization of the objective function is a subject of the DIIS constraints on $c^G$ and $c^\Sigma$. 
Therefore, the corresponding Lagrangian and the Newton step are
\begin{gather}
L^\text{LCIIS}(c^G, c^\Sigma, \lambda) = f(c^G, c^\Sigma) - \lambda^G (1-\sum_i c^G_i) - \lambda^\Sigma (1-\sum_i c^\Sigma_i), \\
\begin{pmatrix}
\mathrm{Re}(H^{GG}) & \mathrm{Re}(H^{G\Sigma}) & \vec{1} & \vec{0}             \\
\mathrm{Re}(H^{\Sigma G}) & \mathrm{Re}(H^{\Sigma \Sigma}) & \vec{0} & \vec{1} \\
\vec{1}^T & \vec{0}^T & 0 & 0 \\
\vec{0}^T & \vec{1}^T & 0 & 0
\end{pmatrix}
\begin{pmatrix}
\Delta c^G \\
\Delta c^\Sigma \\
\lambda^G \\
\lambda^\Sigma 
\end{pmatrix}
= -
\begin{pmatrix}
g^G \\
g^\Sigma \\
0 \\
0
\end{pmatrix}.
\end{gather}
Although the simple system partitioning is not applicable to the system above, 
one can apply a simple diagonal preconditioner to improve the condition number. 
However, such a modification is naive. 
Although it improves the values of the objective function in comparison to LCIIS, 
it does not present a good extrapolation, 
since already for the HF case the coefficients for the Fock matrix and for the density are different 
(which violates the linear connection between the Fock matrix and the density).  

\section{KAIN}
Harrison\cite{Harrison:KAIN:2004} introduced 
the Krylov subspace accelerated inexact Newton method (KAIN).
The goal of KAIN is to find a root $\mathbf{v}_*$ of a nonlinear equation $\mathbf{f(v)} = \mathbf{0}$.
The exact Newton method for the step is given by
\begin{gather}
\mathbf{v}_\text{new} = \mathbf{v} + \Delta \mathbf{v}, \\
\mathbf{F} \Delta \mathbf{v} = - \mathbf{f}, \\
\mathbf{F} = \nabla \mathbf{f},
\end{gather}
where $\mathbf{F}$ is a Jacobian. 
In the subspace version, $i = 1..n$ and the $\mathbf{f}(\mathbf{v}_i) = \mathbf{f}_i$ are residuals. 
A Jacobian approximation can be used, giving an inexact Newton algorithm
\begin{gather}
\mathbf{f}(\mathbf{v}_i) = 
\mathbf{f}(\mathbf{v}_n + \mathbf{v}_i - \mathbf{v}_n) \approx 
\mathbf{f}(\mathbf{v}_n) + (\nabla \mathbf{f}(\mathbf{v}_n)) (\mathbf{v}_i - \mathbf{v}_n), \\
\mathbf{F}_n (\mathbf{v}_i - \mathbf{v}_n) \approx \mathbf{f}_i - \mathbf{f}_n.
\end{gather}
If $\mathbf{P}$ is a projector onto an ($n-1$)-dimensional subspace spanned 
by the differences $\mathbf{v}_i-\mathbf{v}_n$,
the equation for the step can be written as
\begin{gather}
\mathbf{F}_n \mathbf{P} \Delta \mathbf{v} + (1-\mathbf{P})\Delta \mathbf{v} = - \mathbf{f}_n \protect\label{eq:KAIN:sub_gen}.
\end{gather}
Projecting Eq.~\ref{eq:KAIN:sub_gen} onto the space of differences, the subspace KAIN equations are obtained:
\begin{gather}
\braket{\mathbf{v}_i - \mathbf{v}_n | \mathbf{F}_n \mathbf{P} \Delta \mathbf{v} + \mathbf{f}_n } = 0, \\
\mathbf{P} \Delta \mathbf{v}_n = \sum_i^{n-1} c_i (\mathbf{v}_i - \mathbf{v}_n), \\
A c = b \protect\label{eq:KAIN:A_system}, \\
A_{ij} = \braket{\mathbf{v}_i - \mathbf{v}_n | \mathbf{f}_i - \mathbf{f}_n}, \\
b_i = -\braket{\mathbf{v}_i - \mathbf{v}_n | \mathbf{f}_n}.
\end{gather}
The component from the external space is found as
\begin{gather}
(1-\mathbf{P}) \Delta \mathbf{v} = - \left(\sum_{i=1}^{n-1} (\mathbf{f}_i - \mathbf{f}_n)c_i + \mathbf{f}_n \right).
\end{gather}
We apply the KAIN algorithm in the same manner as DIIS, taking the self-energy as $\mathbf{v}$ vectors. 
However, not all choices of DIIS residuals can be used in KAIN. 
The commutator residuals are orthogonal to self-energy, leading to an ill-defined system (Eq.~\ref{eq:KAIN:A_system}). 
In this paper, we use the difference residuals, which do not lead to this issue. 

\section{Step restriction}
The problematic DIIS iterations often have large extrapolation coefficients. 
One can interpret the DIIS and LCIIS extrapolation as a given position (from the last iteration) and a step\cite{Harrison:KAIN:2004}:
\begin{gather}
\Sigma_{\text{extr}} = \sum_i c_i \Sigma_i = \Sigma_n + \sum_i \tilde{c}_i \Sigma_i, \\
F_{\text{extr}} = \sum_i c_i F_i = F_n + \sum_i \tilde{c}_i F_i, \\
\sum_i \tilde{c}_i = 0.
\end{gather}
Here, the $F_n$ and $\Sigma_n$ are interpreted as a current position point and 
the remaining term is interpreted as a step. 
In this picture, an overstep can happen if the step size is too large. 
We used a simplified step restriction measure from Ref.\cite{Harrison:KAIN:2004} 
to understand the impact of overstepping.
If the norm of the coefficients $\tilde{c}_i$ is bigger 
than a chosen trust norm $r$, 
we rescale them and perform extrapolation with the scaled coefficients:
\begin{gather}
\tilde{c}_i^\text{new} =
\frac{r}{|\tilde{c}|} \tilde{c}_i.
\end{gather}
In case of KAIN, we measure the norm of the coefficients and restrict the coefficients if the norm exceeds a user-specified threshold in the following way:
\begin{gather}
|\tilde{c}| = |\tilde{c}_\text{in}| + |\tilde{c}_\text{out}|, \\
\tilde{c}_{\text{in},i}^\text{new} =
\frac{r}{|\tilde{c}|} \tilde{c}_{\text{in},i}, \\
\tilde{c}_{\text{out},i}^\text{new} =
\frac{r}{|\tilde{c}|} \tilde{c}_{\text{out},i}, 
\end{gather}
where the coefficients  $\tilde{c}_\text{in}$ and  $\tilde{c}_\text{out}$ are obtained from the expression of the inner and outer components of the step through the $\mathbf{v}$ vectors (self-energies) and $\mathbf{f}$ vectors (differences of self-energies between the subsequent iterations), respectively.

\section{Numerical results and discussion}
\subsection{Computational details}
We used the double-zeta cc-pVDZ basis sets~\cite{Dunning:ccpvxz:1989,Dunning:ccpvxz:LiNaBeMg,Dunning:ccpvxz:Ca}, 
from the EMSL Basis Set Exchange website~\cite{NewBSE,EMSL-paper}. 
The setup for NiO and Si solids was the same as in the previous publications\cite{Iskakov20,Fei:Nevanlinna:2021,Fei:Caratheodoru:2021}: 
the gth-dzvp-molopt-sr basis\cite{GTHBasis} and gth-pbe pseudopotential\cite{GTHPseudo} were used, 
$6\times 6\times 6$ grid was used for Si, 
$2\times 2\times 2$ grid was used for NiO with doubled unit cell along [111] direction 
(to capture the broken-spin antiferromagnetic solution of type II). 
The gth-tzvp-molopt-sr basis, gth-pade pseudopotential, geometry from the Ref.\cite{Freedman:BiVO3:2019}, 
and $2\times 2\times 2$ grid were used for BiVO$_3$, 
for which the initial guess was used from the converged PBE0 calculation. 
The stretched H$_2$ and N$_2$ molecules were chosen with the internuclear H--H and N--N distances of 3.15\AA; 
the stretched H$_8$ cube was chosen with the edge of 3.15\AA.  
The resolution-of-identity approximation was used in the GW calculations~\cite{Iskakov20}.
The one- and two-electron integrals were generated with PySCF program~\cite{PYSCF}.
The intermediate representation was used for the frequency grids~\cite{Yoshimi:IR:2017}.
All the algorithms are implemented in our in-house code. 

\begin{figure}[!h]
  \includegraphics[width=8cm]{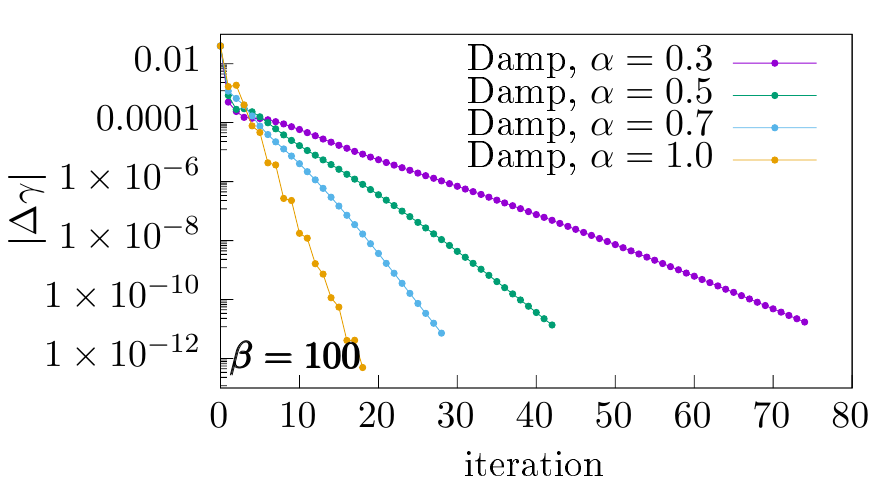}
  \includegraphics[width=8cm]{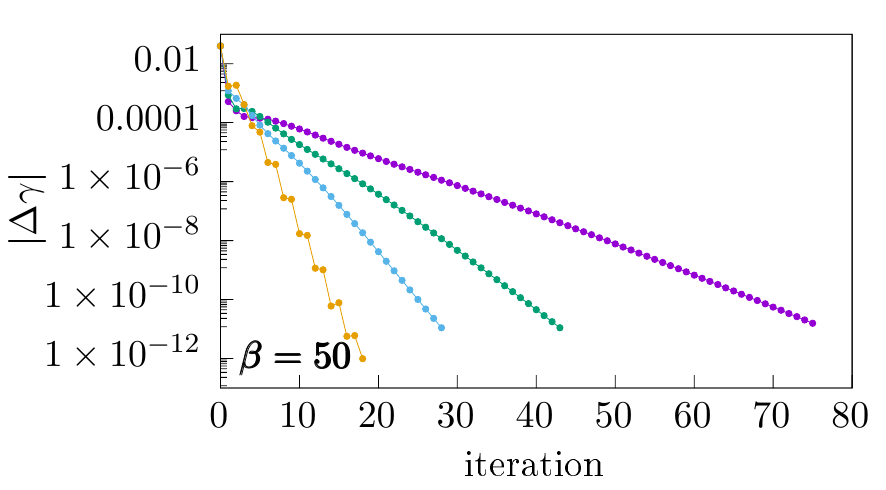} \\
  \includegraphics[width=8cm]{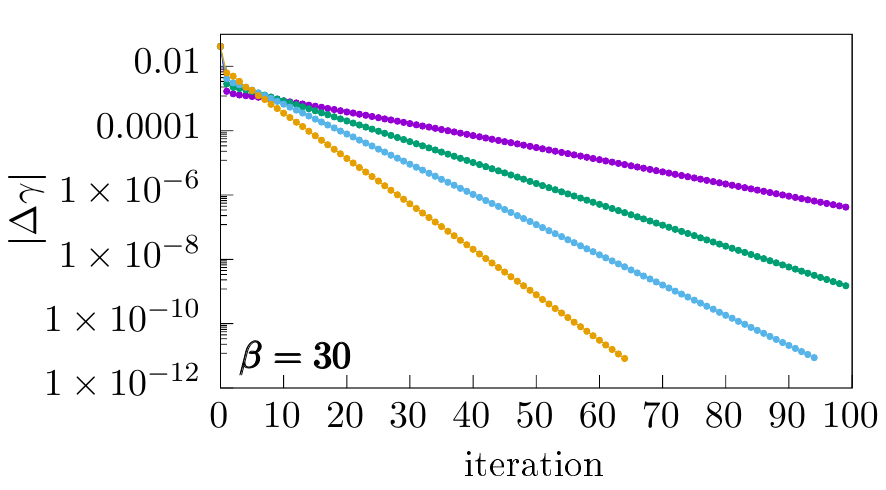} 
  \includegraphics[width=8cm]{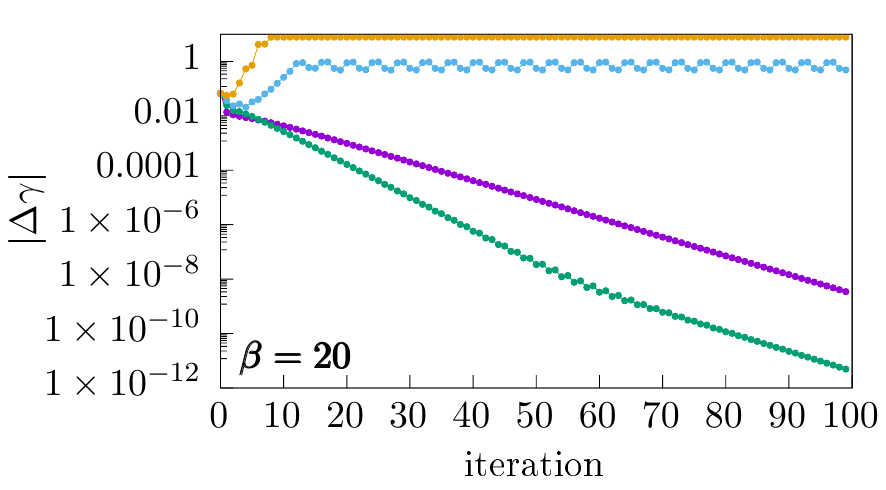} \\
  \includegraphics[width=8cm]{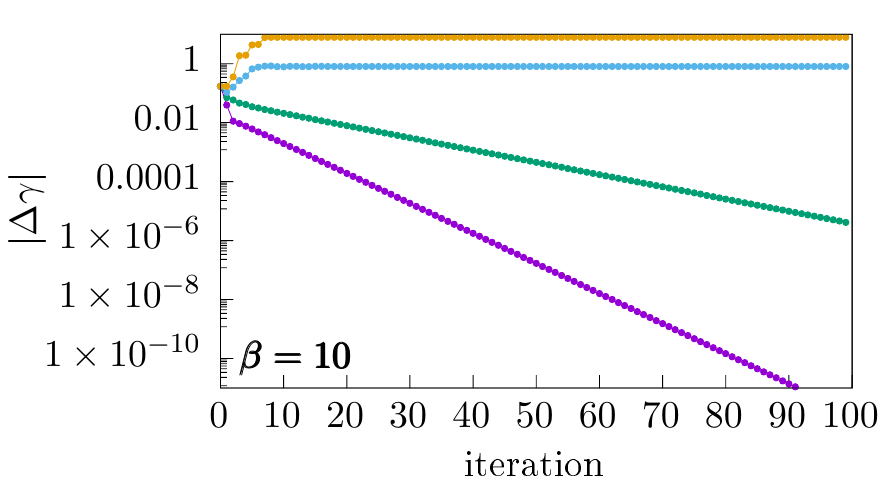}
\centering
\caption{The convergence of damped iterations for GW calculations for a Be atom for a range of inverse temperatures $\beta$. 
         Different values of the damping parameter are shown. 
         \protect\label{fig:Be_damp}}
\end{figure}
\subsection{Atoms}
One of the simplest systems for the investigation of the iterative convergence pattern is a beryllium atom. 
It is a closed-shell system, which does not show 
any problematic behavior in the zero-temperature Hartree--Fock calculations. 
However, the finite-temperature calculations show a different behavior. 
Note that these calculations are performed at a constant chemical potential (not a constant number of particles). 
This can give rise to a varying number of particles during the self-consistent iterations.
In Fig.~\ref{fig:Be_damp}, we plot the convergence of GW iterations with damping 
starting from a zero-temperature HF solution. 
At low temperatures\cite{note:Be_low_T}  ($\beta = 30$--$100$~a.u.$^{-1}$), 
the direct Roothaan-like steps are the fastest. 
However, at higher temperatures ($\beta = 10, 20$~a.u.$^{-1}$), 
the direct steps cause large changes in the average number of electrons (computed from the Green's function at a given iteration), leading to divergence of iterations. 
In this case, a step restriction is necessary. 
More significant damping values ($\alpha = 0.3$ and $0.5$) stabilize the iterations and lead to convergence.

\begin{figure}[!h]
  \includegraphics[width=8cm]{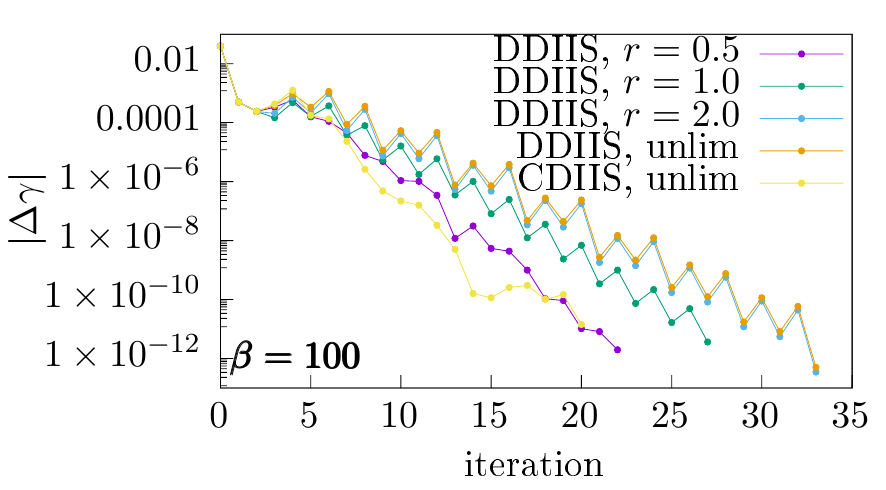}
  \includegraphics[width=8cm]{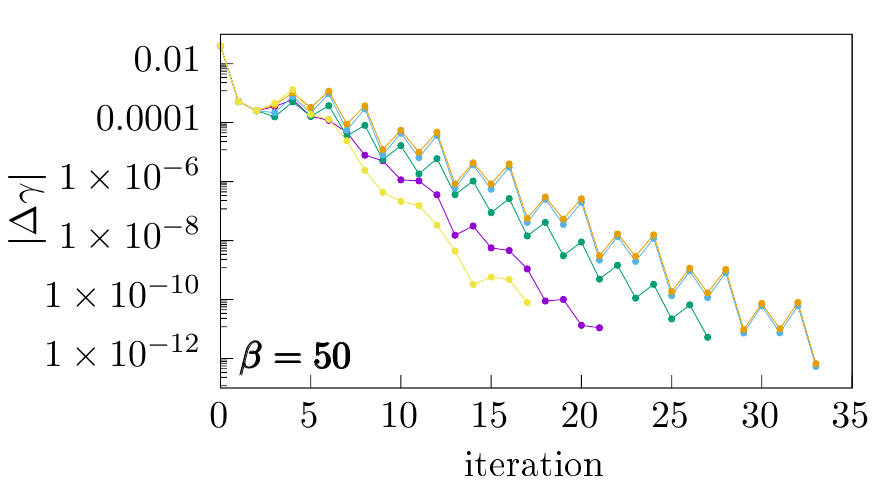} \\
  \includegraphics[width=8cm]{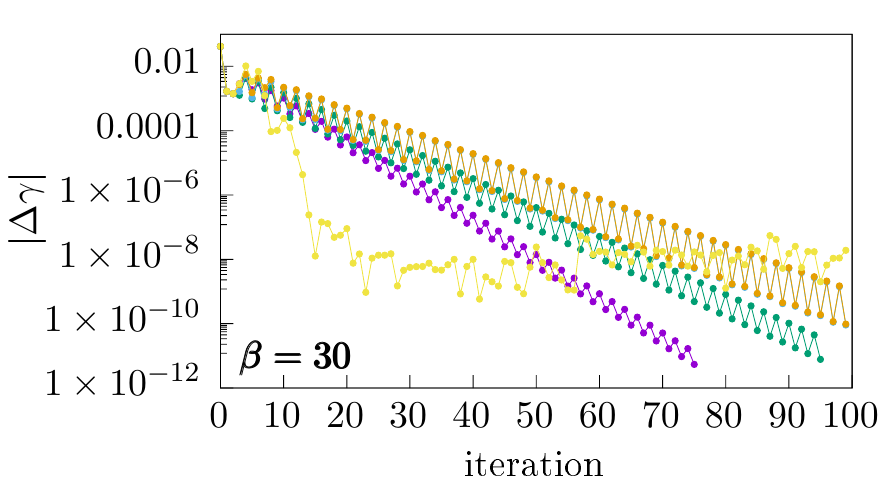}
  \includegraphics[width=8cm]{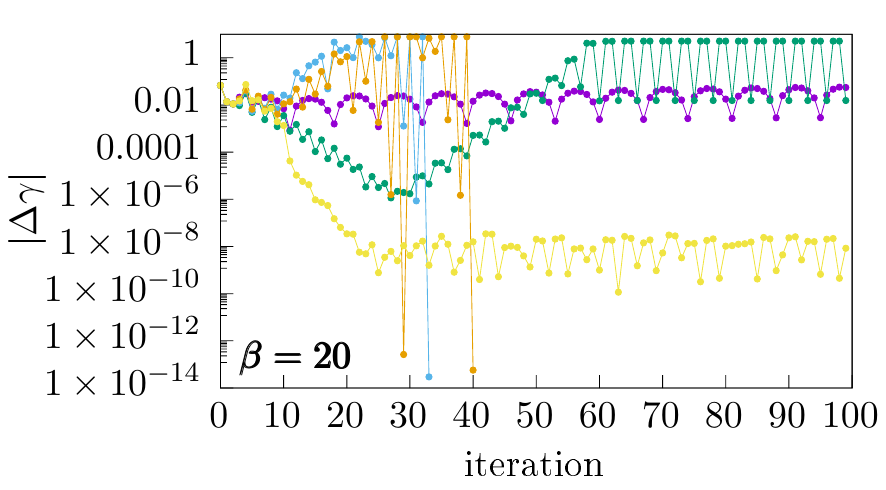}
\centering
\caption{The convergence of DIIS for GW calculations for a Be atom for a range of inverse temperatures $\beta$. 
         Different values of the trust norm $r$ are shown. 
         The calculations where the step restriction has not been applied, are labeled as \emph{unlim}.
         The subspace size is $2$.
         \protect\label{fig:Be_diis}}
\end{figure}
In Fig.~\ref{fig:Be_diis}, we display the DIIS iterations with the difference (DDIIS\cite{note:ddiis}) and 
commutator residuals (CDIIS) 
for the Be atom. 
The DDIIS iterations without step restriction are more oscillatory than the iterations with the step restriction. 
As expected, the smaller the parameter $r$ is, the closer the iterations are to the direct Roothaan steps. 
Yet, even the iterations without restriction perform similar to the slightly damped iterations at a low temperature. 
However, at high temperature, DDIIS with all used values $r$ diverge. 
As shown in SI (Fig~S1), the increase of the subspace size slows the convergence, 
but does not not lead to better stability at high temperature. 
The cause of this behavior is likely the following. 
The DDIIS extrapolation takes into account only self-energy, but not the Green's function, 
which leads to a fluctuation of the number of electrons, 
impacting the computed self-energy from the Green's functions. 
KAIN shows a similar behavior to DDIIS (Fig.~S2). 
This is not accidental, because the step $\Delta \mathbf{v}$ is dominated by the subspace component at most iterations.  
As shown in the Ref.\cite{Harrison:KAIN:2004}, this regime is expected to be similar to DIIS. 
Fig.~S2 shows that the divergence patterns at high temperature are similar to DIIS as well. 
The unsatisfactory behavior of the difference residual has been reported before 
in the context of coupled-cluster equations, 
where it also led to oscillatory behavior\cite{Matthews:microiter:Lambda:2020}.

Commutator residuals, however, lead to a much more stable behavior of iterations. 
This is not surprising since the commutator residuals include both the Green's functions and self-energies. 
Fig~S3 shows commutator residuals within DIIS. 
The CDIIS converges very rapidly with a loose convergence criterion 
As the changes in the density matrix and energy become of the order of $10^{-7}$ a.u., 
the iterations oscillate and do not converge to a tight convergence. 
The likely reason is in the numerical errors in the commutator residual evaluation.  
Moreover, this also makes the extrapolation coefficients sensitive to the numerical noise 
coming from the evaluation of the commutators leading to a noisy estimate of the error and noisy extrapolation coefficients, 
preventing convergence with a tight convergence criterion. 

\begin{figure}[!h]
  \includegraphics[width=8cm]{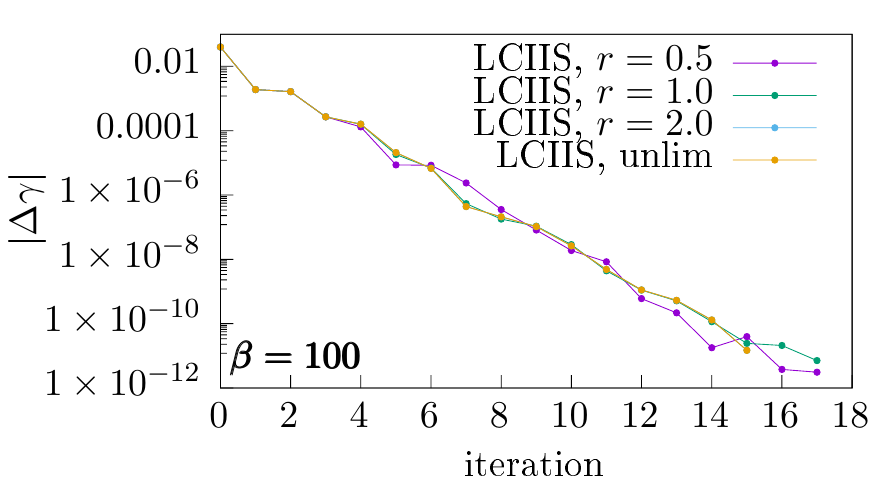}
  \includegraphics[width=8cm]{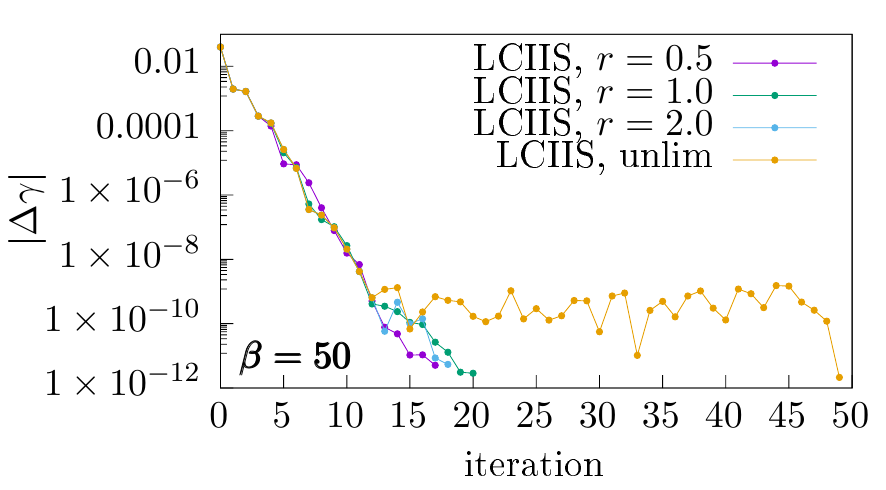} \\
  \includegraphics[width=8cm]{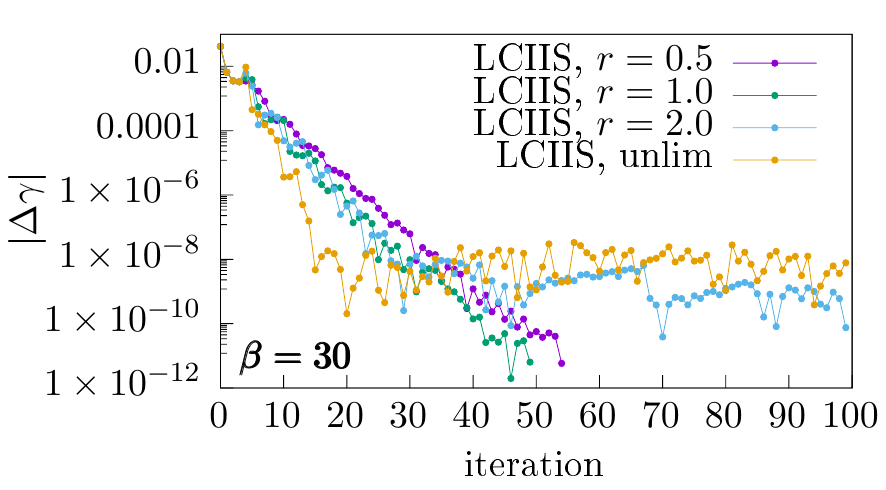}
  \includegraphics[width=8cm]{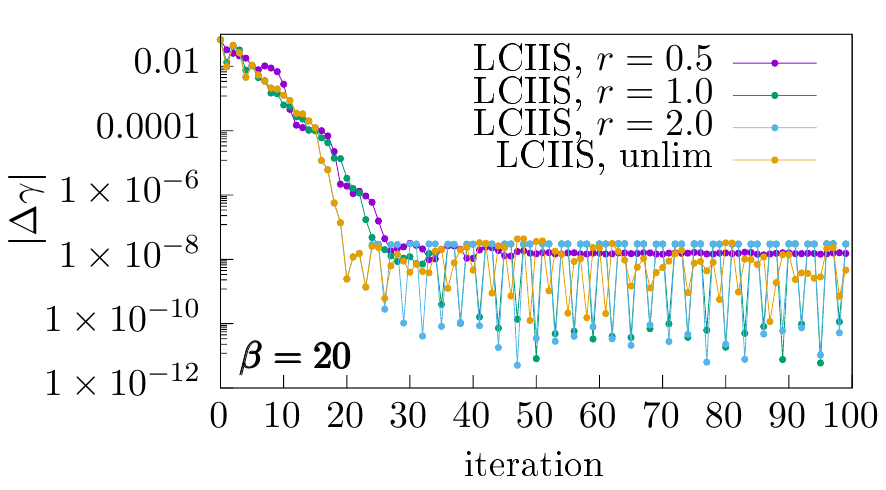}
  \includegraphics[width=8cm]{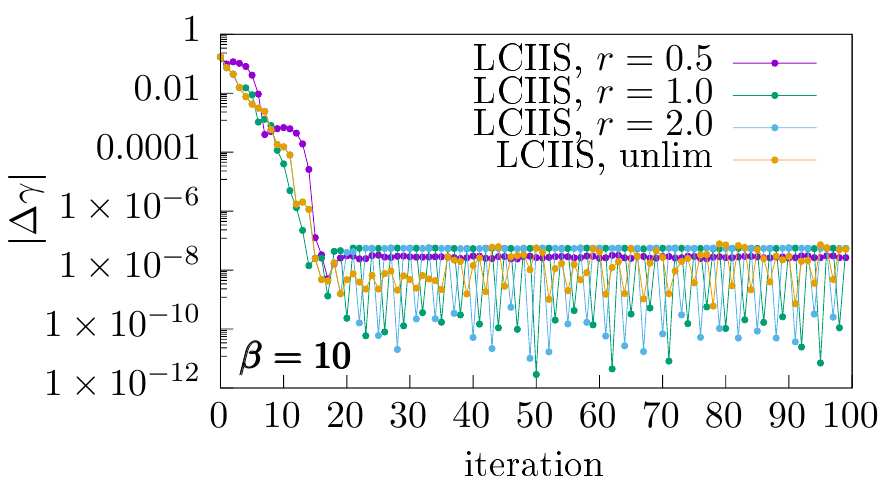}
\centering
\caption{The convergence of LCIIS for GW calculations for a Be atom for a range of inverse temperatures $\beta$. 
         Different values of the trust norm $r$ are shown. 
         The calculations where the step restriction has not been applied, are labeled as \emph{unlim}.
         The subspace size is $2$.
         \protect\label{fig:Be_lciis}}
\end{figure}

LCIIS (Fig.~\ref{fig:Be_lciis}) behaves similarly to CDIIS. 
The trust norm does not have a strong impact on the convergence of iterations, 
partially suppressing fluctuations near convergence (the same is true for CDIIS). 
We recommend to switch to a stable damping near convergence if the tight convergence is required for LCIIS and CDIIS. 

The Fig.~S4 and S5 in SI show the convergence of iterations for Mg and Ca atoms 
at different temperatures, where the chemical potential is optimized after each extrapolation 
to yield a number of electrons for neutral systems. 
The best performing algorithms are CDIIS and LCIIS, the same as for the Be atom.

\subsection{Solids}
\begin{figure}[!h]
  \includegraphics[width=8cm]{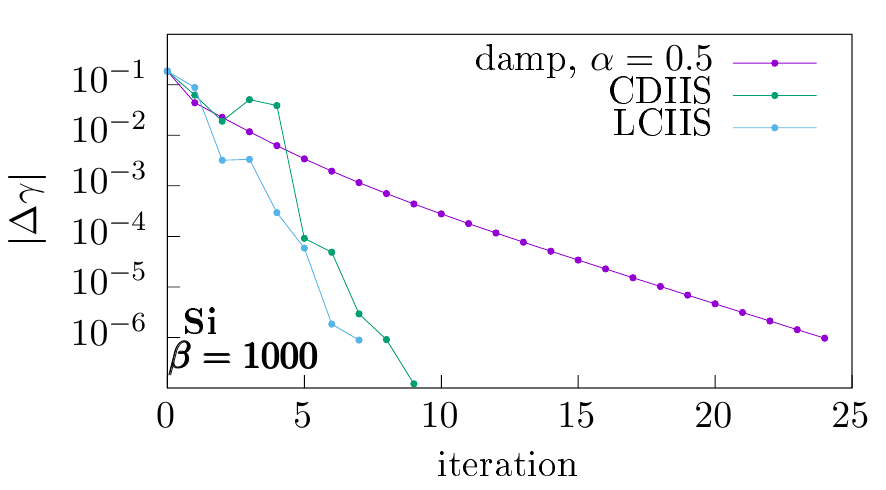}
  \includegraphics[width=8cm]{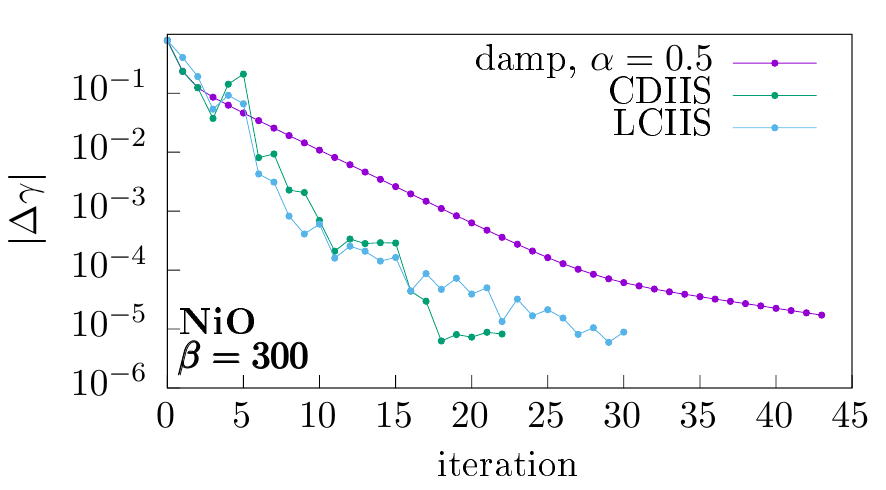} \\
  \includegraphics[width=8cm]{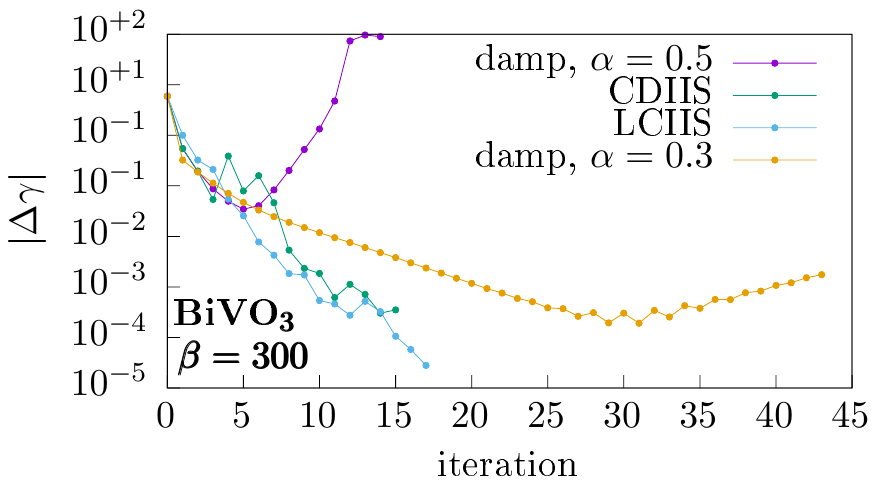} 
\centering
\caption{The convergence of CDIIS and LCIIS for GW calculations for solids: 
         Si (left top), NiO (right top), and BiVO$_3$ (bottom).
         The subspace size is $3$ for Si and NiO and $5$ for BiVO$_3$.
         \protect\label{fig:solids}}
\end{figure}
Fig.~\ref{fig:solids} shows the performance of CDIIS and LCIIS for solid Si, NiO, and BiVO$_3$. 
CDIIS and LCIIS show a similar performance, both significantly outperforming simple damping. 
Interestingly, for NiO, the convergence of iterations with constant damping slows down when $|\Delta \gamma|$ reaches small values. 
For BiVO$_3$, we even observe divergences when constant damping is used. This is probably due to complicated electronic structure of these solids
that includes both covalent and ionic contributions. 
Note that for both NiO and BiVO$_3$ with simple damping ($\alpha = 0.5$ and $0.3$), the convergence of the energy ($10^{-7}$ and $10^{-6}$~a.u., respectively) is achieved, however, $|\Delta \gamma|$ is either showing a slow down in convergence or even a diverging pattern.
This highlights that the convergence of the density matrix should be checked in addition to the usual convergence criteria based on energy.

\subsection{Difficult cases}
\begin{figure}[!h]
  \includegraphics[width=8cm]{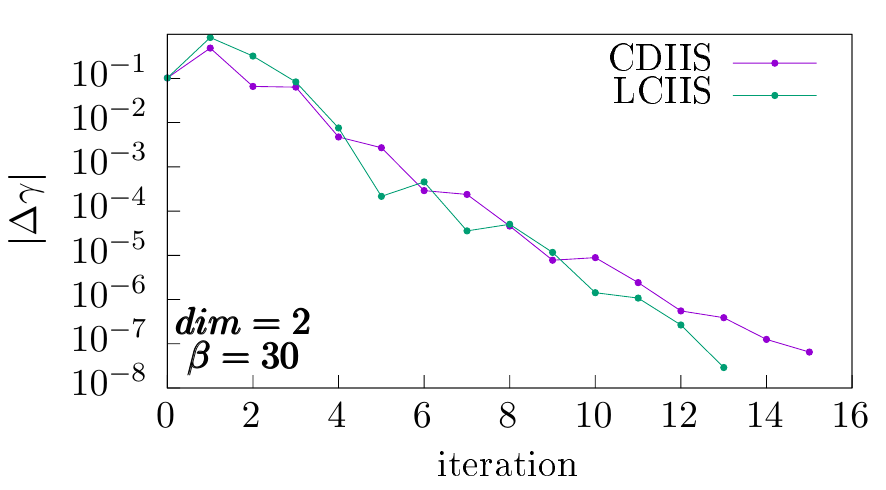}
  \includegraphics[width=8cm]{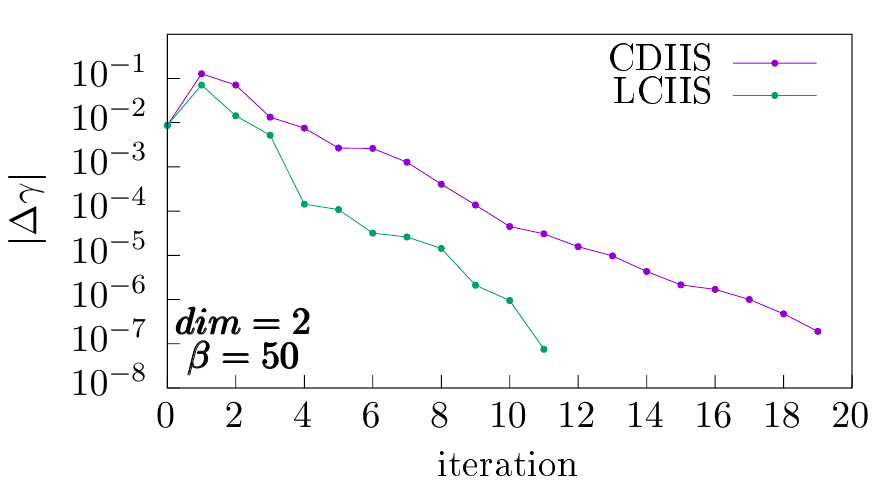} \\
  \includegraphics[width=8cm]{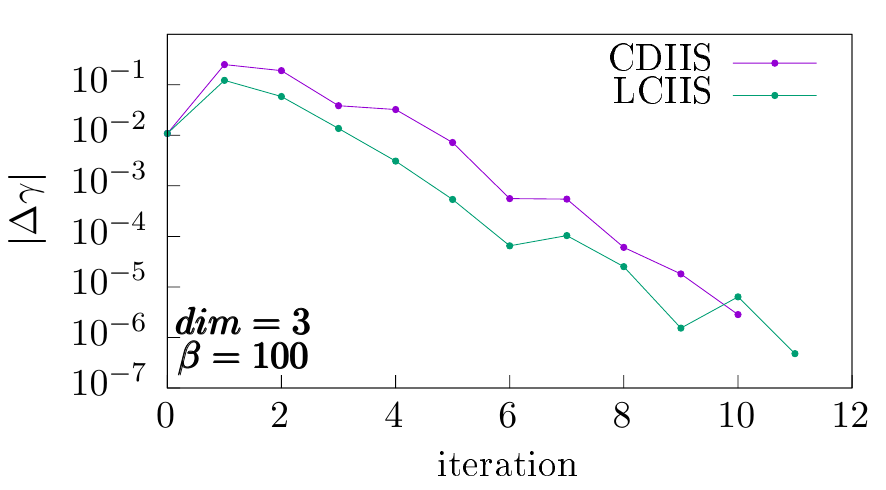}
  \includegraphics[width=8cm]{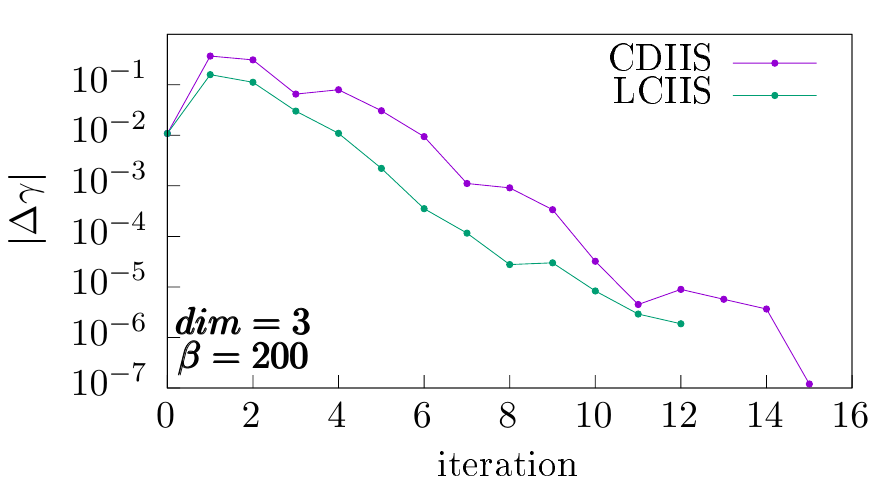} \\
  \includegraphics[width=8cm]{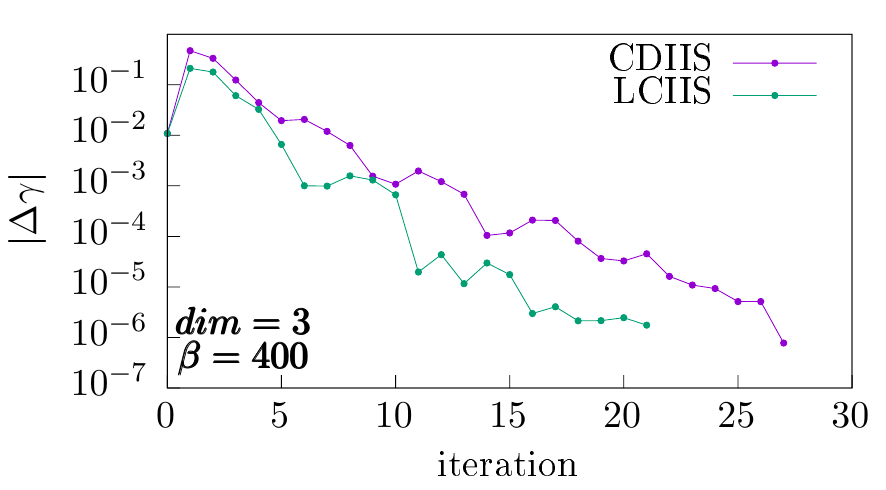}
  \includegraphics[width=8cm]{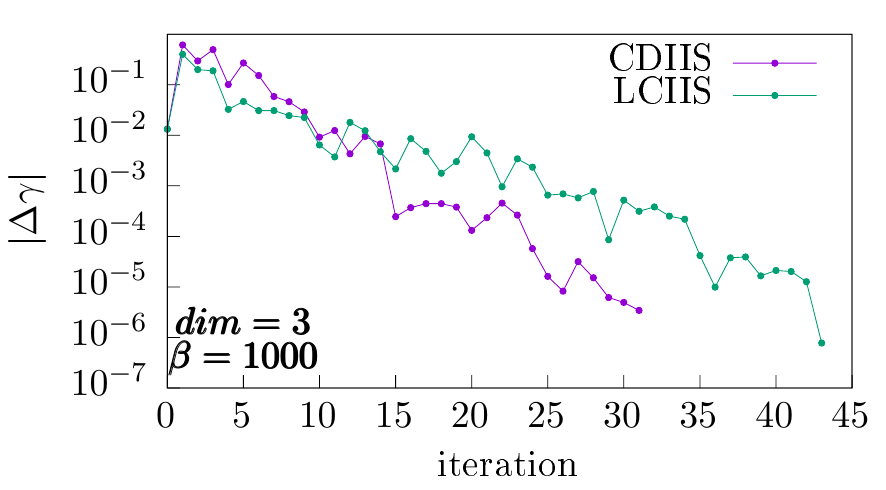} 
\centering
\caption{The convergence of CDIIS and LCIIS for RGF2 calculations for a stretched H$_2$ molecule for different temperatures. 
         The initial guesses for each run are taken either from the finite-temperature HF ($\beta=30$) 
         or from the RGF2 at a higher temperature (all other graphs).
         The subspace size is indicates in each of the graphs as \emph{dim}.
         \protect\label{fig:H2}}
\end{figure}

For low-order methods such as GW or GF2 achieving convergence in strongly correlated molecules or solids 
is notoriously difficult. 
For example, the restricted GF2 calculation at a low temperature ($\beta=1000$~a.u.$^{-1}$) for the stretched H$_2$ molecule starting from a zero-temperature RHF guess 
fails to converge even if a severe damping is used. 
The oscillations between the iterations are so large that they cannot be efficiently damped, 
and both CDIIS and LCIIS algorithms do not converge (Fig~S6 in SI). 
We found that for such cases a better starting guess is crucial for achieving convergence. 
At a high temperature ($\beta=30$~a.u.$^{-1}$ for stretched molecules studied here), 
the strong correlation between nearly degenerate orbitals is effectively diminished\cite{note:high_T} 
and RGF2 calculation converges rapidly when started from a high-temperature HF guess 
(Fig.~\ref{fig:H2} shows such convergence for stretched H$_2$).
Here, we subsequently used the converged self-energies and Fock matrices form the higher temperature calculations as an initial guess for the lower temperature calculations (leading to gradual cooling of the system).  
Fig.~S7 in SI shows how the RGF2 converges with CDIIS and LCIIS for a stretched H$_8$ cube. 
The convergence pattern is nearly the same as for the stretched H$_2$, 
and the increase in the number of strongly correlated 
radical centers does not worsen the convergence. 
Similarly to the known practice of CDIIS in the zero-temperature SCF calculations, 
the increase of the subspace size stabilizes iterations. 
While the subspace size of 2 is sufficient to converge RGF2 for stretched H$_2$ and H$_8$ at high temperature, 
one needs to increase the subspace to 3 in order to converge calculations at lower temperatures. 
This strategy can be use to converge the stretched N$_2$ as well (Fig.~S8 in SI), 
but it requires a large extrapolation subspace (dim $ = 5$) to stabilize iterations. 
This is not surprising since breaking a triple bond leads to a stronger correlation than breaking a single bond. 

\section{Conclusions}
We presented an application of subspace convergence acceleration algorithms to iterative Green's function methods containing the Dyson equation. 
For the convergence acceleration algorithms analyzed here, the introduced computational overhead is negligible in comparison with the cost of the evaluation of self-energy during iterations, 
making the approach very promising for treating both molecules and solids. 

We used the difference self-energy residuals for DIIS and KAIN. 
We generalized the concept of commutator residuals to an arbitrary treatment of electron correlation 
via a self-energy functional and used it to generalize CDIIS and LCIIS. 
On a number of examples, we showed that the choice of residuals is crucial. 
If the chemical potential is fixed, the difference self-energy residuals often lead to divergencies of iterations (due to fluctuations in the average number of electrons), 
even if the overstep treatment is applied. 
The commutator residuals with CDIIS and LCIIS behave much more regularly, 
both with the chemical potential being fixed and optimized. 

For molecules and solids, for the cases tested, we observed that CDIIS and LCIIS performance is comparable.
Both these algorithms outperform simple damping in almost all the cases. 
If the high-temperature initial guess is given, CDIIS and LCIIS dramatically extend the applicability of the restricted Green's function methods 
to strongly correlated systems, (such as stretched H$_2$, H$_8$, and N$_2$ examined here), where achieving convergence was either impossible or very difficult before.

\section*{Acknowledgments}
 P.P. and D.Z. acknowledge support from NSF grant CHE-1453894.
Ch-N. Y. acknowledges support of the Center for Scalable, Predictive methods for Excitation and Correlated phenomena (SPEC), which is funded by the U.S. Department of Energy (DOE), Office of Science, Office of Basic Energy Sciences, the Division of Chemical Sciences, Geosciences, and Biosciences. We thank Dr. Robert J. Harrison for his comment regarding the use of commutator residuals with KAIN algorithm. 
We thank Yanbing Zhou for providing the input data for the BiVO$_3$ system.

\section*{Supplementary Material}
Iterations of DDIIS, CDIIS, KAIN, LCIIS for atoms and molecules. 

\renewcommand{\baselinestretch}{1.5}

\clearpage
\bibliographystyle{prf}

\begin{thebibliography}{10}
\bibitem{note:DFT_lin}
This is true only for Hartree--Fock. In DFT, the Kohn--Shan matrix does not
  depend on density linearly, but direct application of the LCIIS still gives
  good performance.
\end{thebibliography}

\end{document}